\documentclass[conference]{IEEEtran}
\IEEEoverridecommandlockouts
\usepackage{amsmath}
\usepackage{amsfonts}
\usepackage{bbding}
\usepackage{amssymb}
\usepackage{array}
\usepackage{subfigure}

\usepackage{graphicx}
\usepackage{subfigure}
\usepackage[named]{algo}
\usepackage{algorithmic}
\usepackage{psfrag}
\usepackage{stfloats}
\usepackage[compress]{cite}
\makeatletter
\renewcommand{\citepunct}{,\penalty\@m\hskip.13emplus.1emminus.1em}
\renewcommand{\citedash}{\hbox{--}\penalty\@m}
\makeatother
\usepackage{setspace}
\usepackage{color}
\allowdisplaybreaks

\usepackage{amsthm}
\usepackage{stfloats}

\usepackage{hyperref}
\usepackage{bm}

\begin{document}
\title{Cross-layer Transmission Design for Tactile Internet}

\author{
\IEEEauthorblockN{{Changyang She and Chenyang Yang}} \vspace{0.0cm}
\IEEEauthorblockA{School of Electronics and Information
Engineering,\\ Beihang University, Beijing, China\\
Email:  \{cyshe,cyyang\}@buaa.edu.cn}\vspace{-0.6cm}\thanks{This work is supported partially by the National High Technology Research and Development Program of China under grant 2014AA01A703, National Basic Research Program of China, 973 Program under grant 2012CB316003 and National Natural Science Foundation of China (NSFC) under Grant 61120106002.}
\and \IEEEauthorblockN{{Tony Q. S. Quek}} \vspace{0.0cm}
\IEEEauthorblockA{$\quad\quad$Information Systems Technology and Design Pillar,\\ $\quad\quad\quad\quad$Singapore University of Technology and Design, Singapore \\
Email: tonyquek@sutd.edu.sg} }

\maketitle
\begin{abstract}
To ensure the low end-to-end (E2E) delay for tactile internet, short frame structures will be used in 5G systems. As such, transmission errors with finite blocklength channel codes should be considered to guarantee the high reliability requirement. In this paper, we study cross-layer transmission optimization for tactile internet, where both queueing delay and transmission delay are accounted for in the E2E delay, and different packet loss/error probabilities are considered to characterize the reliability. We show that the required transmit power becomes unbounded when the allowed maximal queueing delay is shorter than the channel coherence time. To satisfy quality-of-service requirement with finite transmit power, we introduce a proactive packet dropping mechanism, and optimize a queue state information and channel state information dependent transmission policy. Since the resource and policy for transmission and the packet dropping policy are related to the \emph{packet error probability}, \emph{queueing delay violation probability}, and \emph{packet dropping probability}, we optimize the three probabilities and obtain the policies related to these probabilities. We start from single-user scenario and then extend our framework to the multi-user scenario. Simulation results show that the optimized three probabilities are in the same order of magnitude. Therefore, we have to take into account all these factors when we design systems for tactile internet applications.
\end{abstract}

\section{Introduction}
{Tactile internet} enables unprecedented mobile applications such as autonomous vehicles, mobile robots, augmented reality and factory automation\cite{Meryem2016Tactile}, which calls for ultra-low latency (say 1 ms) and ultra-high reliability (say $99.99999$\%). In the fifth generation (5G) communication systems, achieving such an extremely stringent quality-of-service (QoS) has become one of the major goals \cite{A2014Scenarios}. Ensuring the short end-to-end (E2E) delay and low packet loss/error probability calls for new air interface for 5G systems. By introducing short frame structure and short transmit time interval (TTI), transmission delay can be reduced \cite{Petteri2015A}. With short frame, the channel coding is then performed with a finite block of symbols under ultra-high reliability requirement, and hence the transmission error probability should be considered. Compared with the channel capacity, the maximal achievable rate with finite blocklength channel codes under given transmission error probability requirement is more relevant to our current problem \cite{Yury2014Quasi}.

Though important, the queueing delay is largely overlooked in most of the existing literatures that study ultra-short delay and ultra-high reliability transmissions. For applications with medium delay requirement, the throughput with finite blocklength codes under statistical queueing constraint was studied in \cite{Throughput2011Mustafa}, where the delay is much larger than the channel coherence time. However, in tactile internet applications, the ultra-short delay could be shorter than the coherence time. When the average delay approaches the coherence time, the average transmit power may become unbounded  \cite{Berry2013}.

In this paper, we study cross-layer  optimization for tactile internet. To ensure the QoS requirement of ultra-low E2E delay and ultra-high reliability, the transmission delay and error probability as well as the statistical queueing delay requirement (characterized by a delay bound and a small delay violation probability) are considered.  To satisfy the QoS with finite transmit power, a proactive packet dropping mechanism is introduced. Since the packet error
probability, queueing delay violation probability, and packet
dropping probability depend on the transmission resource and policy as well as the packet dropping policy, we optimize them together to minimize the transmit power of the base station (BS). We first optimize the power allocation and packet dropping policy in each TTI in single-user scenario, then extend our framework to multi-user scenario and optimize the bandwidth allocation among users. Simulation results show that the optimized three probabilities are in the same order of magnitude.

\section{System Model}
\begin{figure}[htbp]
        \vspace{-0.0cm}
        \centering
        \begin{minipage}[t]{0.38\textwidth}
        \includegraphics[width=1\textwidth]{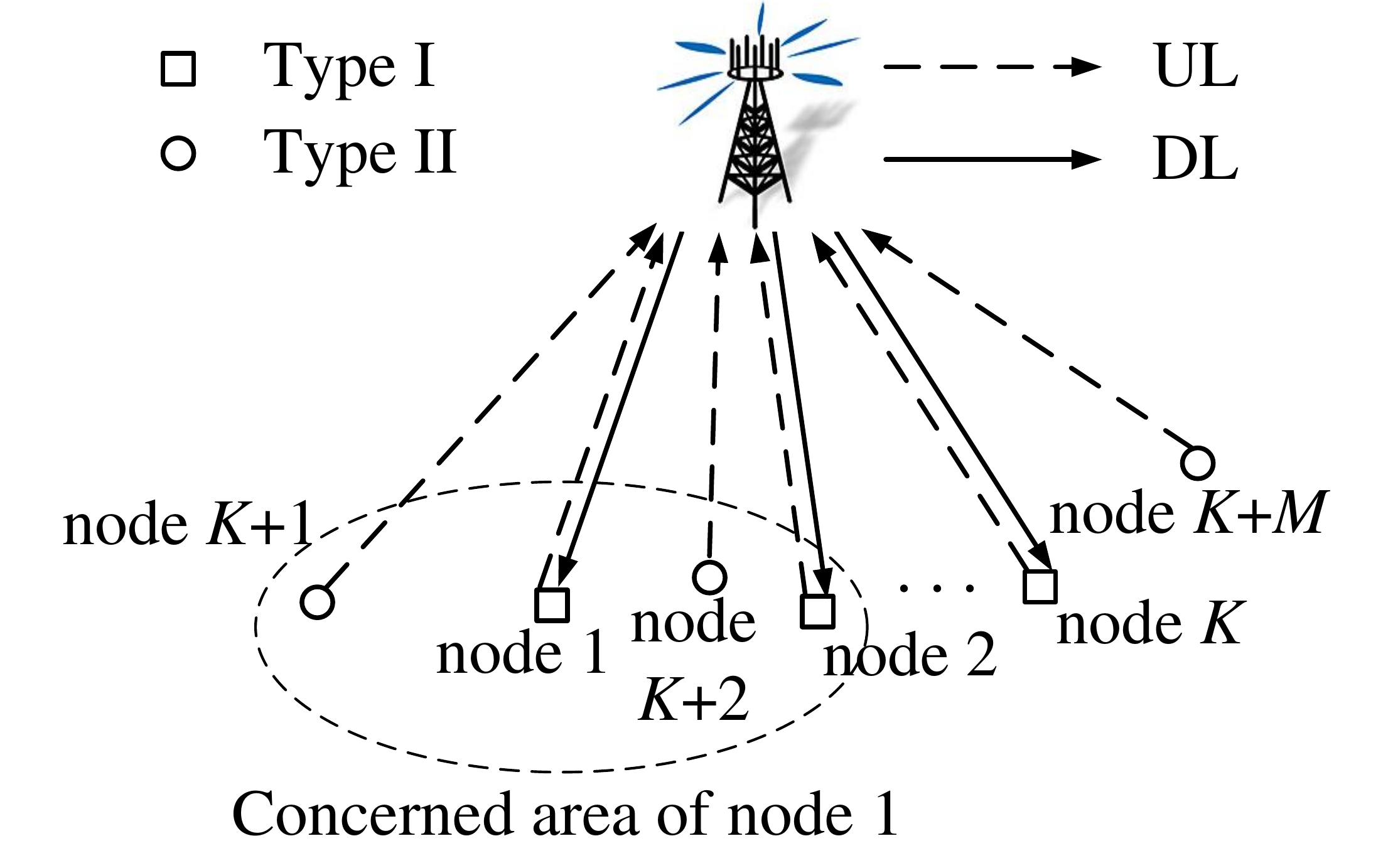}
        \end{minipage}
        \vspace{-0.3cm}
        \caption{System model.}
        \label{fig:smodel}
        \vspace{-0.4cm}
\end{figure}

Consider a time division duplexing cellular system, which consists of a BS with $N_\mathrm{t}$ antennas and $K+M$ single-antenna nodes, as shown in Fig. \ref{fig:smodel}. The nodes are divided into two types. The first type of nodes are $K$ users, which need to upload and download packets to the BS. The second type of nodes are $M$ sensors, which only upload packets. Time is discretized into frames. As illustrated in Fig. \ref{Illustration}, each frame has duration $T_\mathrm{f}$ and consists of a downlink (DL) transmission phase with duration $\phi$ and an uplink (UL) transmission phase with duration $\varphi$. In the UL phase, all the nodes upload their own messages in short packets to the BS. In the DL phase, the BS processes the UL received messages from the nodes that lie in the concerned area of each user, and then transmits the relevant messages to the target users. Since the interference among nodes cause severe deterioration in QoS, we assume a frequency division multiple access system.

\vspace{-0.2cm}
\subsection{Reliability and Delay Metrics}
For tactile service, the QoS can be characterized by a maximal E2E delay for each packet,  $D_{\max}$, and a maximal packet loss/error probability, ${\varepsilon _\mathrm{D}}$. The E2E delay is very short, say 1 ms \cite{Gerhard2014The}, which includes UL and DL transmission delay and queueing delay in the buffer of BS. Denote the size of each packet as $u$~bits. We assume $u$ is small enough such that it can be transmitted within one UL phase. Then, we focus on the DL transmission and investigate how to achieve the stringent QoS requirement.

To ensure ultra-low transmission delay, we consider short frame structure proposed  in \cite{Petteri2015A}, where the TTI is the same as the frame duration and  $T_\mathrm{f} \ll D_{\max}$. Moreover, we assume that the DL transmission can be finished within the duration of $\phi$. Then, the queueing delay for every packet should be bounded as $D^q_{\max} \triangleq D_{\max}-T_\mathrm{f}$ with a small violation probability ${\varepsilon_k^q}$. With finite blocklength channel codes, the transmission of each packet can be finished within one frame with a small error probability ${\varepsilon_k^c}$.  As detailed later, to ensure the statistical requirement imposed on the queueing delay for each packet $(D^q_{\max}, {\varepsilon_k^q})$, the required  transmit power may become unbounded in deep fading. To guarantee the E2E delay and reliability with finite transmit power, we proactively discard some packets from the head of the queue under deep fading, and control the overall E2E reliability as follows:
\begin{align}
&1 - (1- {\varepsilon_k^c})(1- {\varepsilon_k^q})(1- {\varepsilon_k^h})\approx {\varepsilon_k^c} + {\varepsilon_k^q} +{\varepsilon_k^h} \leq {\varepsilon _\mathrm{D}},\label{eq:reliability}
\end{align}
where ${\varepsilon_k^h}$ is the proactive packet dropping probability for the $k$th user. user. Note that the above approximation is valid since ${\varepsilon_k^c}$, ${\varepsilon_k^q}$ and ${\varepsilon_k^h}$ are extremely small.

\subsection{Channel Model}
\begin{figure}[htbp]
        \vspace{-0.3cm}
        \centering
        \begin{minipage}[t]{0.45\textwidth}
        \includegraphics[width=1\textwidth]{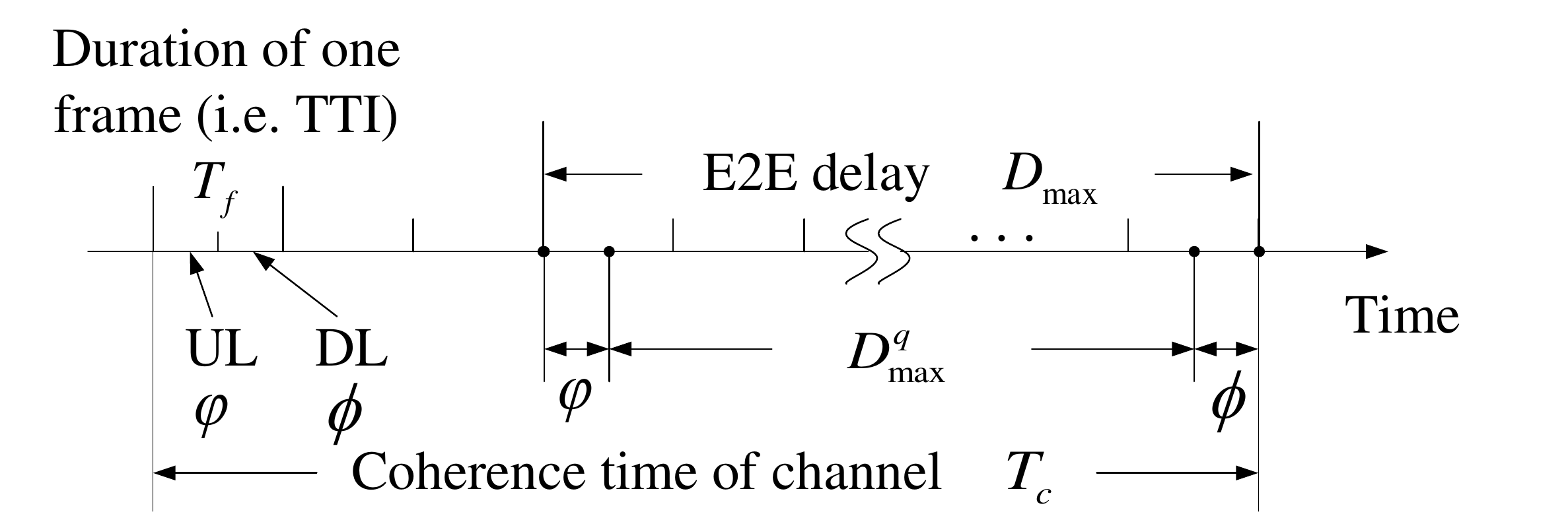}
        \end{minipage}
        \vspace{-0.2cm}
        \caption{Illustration of TTI, E2E delay and channel coherence time.}
        \label{Illustration}
        \vspace{-0.4cm}
\end{figure}

Denote the coherence time of the channel as $T_{\rm c}$. Since the E2E delay is extremely short, we assume that $T_{\rm c} > D_{\max} > D^q_{\max}$ as illustrated in Fig. \ref{Illustration}.\footnote{For instance, for users with velocities less than $120$~km/h and the system operating in carrier frequency of $2$~GHz, the channel coherence time is larger than $1$~ms, which exceeds the delay bound of each packet.}

When the number of users is large or the overall bandwidth is small, each user will be allocated with bandwidth that is smaller than the coherence bandwidth of the channel. When channel coding is performed within each frame, which is shorter than the coherence time, the channel is referred to as \emph{quasi-static fading channel} \cite{Yury2014Quasi}. Denote the average channel gain and channel vector of the $k$th user in a certain coherence block as $\alpha_k$ and ${\bf{h}}_k \in {\mathbb{C}}^{N_\mathrm{t} \times 1}$, whose elements are independent and identically complex Gaussian distributed with zero mean and unit variance. According to the {normal approximation} in \cite{Yury2014Quasi}, when $\alpha_k$ and  ${\bf{h}}_k$ are perfectly known at the BS and the user, the maximal number of packets that \emph{can be} transmitted to the $k$th user in the $n$th frame (i.e., in the $n$th TTI) can be accurately approximated as
\begin{align}
s_k(n) \approx \frac{ \phi W_k}{u \ln{2}} \left\{\ln\left[1+\frac{\alpha_k P _k(n)g_k}{ N_0 W_k}\right] - \sqrt{\frac{V}{\phi W_k}}f_{\rm G}^{-1}({\varepsilon_k^c})\right\}, \label{eq:sn}
\end{align}
where $W_k$ is the bandwidth allocated to the $k$th user, $P _k(n)$ is the transmit power allocated to the $k$th user according to its queue length and channel state in the $n$th frame, $g_k = {\bf{h}}_k^H{\bf{h}}_k$, $[ \cdot ]^H$ denotes the conjugate transpose, $N_0$ is the single-sided noise spectral density, $f_{\rm G}^{-1}(x)$ is the inverse of the Gaussian Q-function,  and $V$ is the channel dispersion given by \cite{Yury2014Quasi}
\begin{align}\label{eq:dispersion}
V = 1-\frac{1}{\left[1+\frac{\alpha_k P _k(n)g_k}{ N_0 W_k}\right]^2}.
\end{align}
Then, the number of symbols transmitted in the DL transmission phase of one frame, $n^s_k$, is determined by the bandwidth and transmission time, i.e., $n^s_k = \phi W_k$.

\subsection{Queue Model}
In the $n$th frame, the $k$th user requests packets that are uploaded from its nearby nodes. The indices of the nodes that lie in the concerned area of the $k$th user constitute a set ${\mathcal{A}}_k$ with cardinality $\left|{\mathcal{A}}_k\right|$. As illustrated in Fig. \ref{fig:smodel}, the index set of the nearby nodes of node $1$ is ${\mathcal{A}}_1 = \{2,K+1,K+2\}$.
Then, the number of packets waiting in the queue for the $k$th user at the beginning of the $(n+1)$th frame can be expressed as
\begin{align}
Q_k\left( {n + 1} \right) = \max \left\{ {Q_k\left( n \right) - s_k\left( n \right)  },0 \right\}+ \sum\limits_{i \in {\mathcal{A}}_k} {{a_i}\left( n \right)}, \label{eq:queue}
\end{align}
where $a_i\left( n \right)$, $i \in {\mathcal{A}}_k$ is the number of packets uploaded to the BS from the $i$th nearby node of the $k$th user. Denote the number of packets departed from the $k$th queue in the $n$th frame as $b_k(n)$. If all the packets in the queue can be completely transmitted in the $n$th frame, then $b_k(n) = Q_k(n)$. Otherwise, $b_k(n)=s_k(n)$. Hence, we have
\begin{align}
b_k(n) = \min \left\{Q_k\left( n \right), s_k\left( n \right)\right\}. \label{eq:bn}
\end{align}

\section{Ensuring the QoS Requirement}
In this section, we first employ {effective bandwidth}, a widely used design tool \cite{EB}, to represent the statistical queueing delay requirement. Then, we show that the required transmit power to ensure $(D^q_{\max}, {\varepsilon_k^q})$  for some packets may become unbounded. To guarantee the QoS on $D_{\max}$ and ${\varepsilon_\mathrm{D}}$ with finite transmit power, we propose a proactive packet dropping mechanism. For notational simplicity, we consider a single-user scenario herein, where $K=1$ and $M > 1$. As such, the index $k$ can be omitted.

\subsection{Queueing Delay Requirement}
For stationary packet arrival process $\{{\sum\limits_{i \in {\mathcal{A}}}{{a_i}\left( n \right)} }, n = 1,2,...\}$, the effective bandwidth is defined as \cite{EB}
\begin{align}
E^{B}(\theta)
&= \mathop {\lim }\limits_{N \to \infty } \frac{1}{{NT_\mathrm{f}{\theta }}}\ln \left\{ {{\mathbb{E}}\left[ {\exp \left( {{\theta}\sum\limits_{n = 1}^N {\sum\limits_{i \in {{\cal A}}} {{a_i}\left( n \right)} } } \right)} \right]} \right\},\label{eq:EB}
\end{align}
where $N$ is the number of frames, and $\theta$ is the QoS exponent. A large value of $\theta$ indicates a stringent delay requirement. According to \cite{Tang2007Quality}, when the delay bound approaches the coherence time, the power allocation over fading channel is channel inversion. In other words, the service rate becomes a constant. When the constant service rate equals to $E^{B}(\theta)$, the steady state queueing delay bound violation probability can be approximated as \cite{EC}
\begin{align}
\Pr\{D(\infty) > D^q_{\max}\} \approx \eta \exp\{-\theta E^B(\theta) D^q_{\max}\},\label{eq:apporxD}
\end{align}
where $\eta$ is the buffer non-empty probability. Since $\eta\leq 1$, we have
\begin{align}
\Pr\{D(\infty) > D^q_{\max}\} \leq \exp\{-\theta E^B(\theta) D^q_{\max}\} \triangleq P_\mathrm{D}^{\rm UB}. \label{eq:UB}
\end{align}
If the upper bound in \eqref{eq:UB} satisfies
\begin{align}
P_\mathrm{D}^{\rm UB} = \exp\{-\theta E^B(\theta) D^q_{\max}\} = \varepsilon^q, \label{eq:delay}
\end{align}
then the requirement $(D^q_{\max}, \varepsilon^q)$ can be satisfied.

{\bf{Remark 1:}} Note that \eqref{eq:UB} is not a strict upper bound for all kinds of arrival processes since \eqref{eq:apporxD} is accurate when the queue length is large enough. However, \eqref{eq:UB} is still a valid upper bound for Poisson arrival process even in very short delay regime \cite{Changyang2016TVC}. If the packet arrival process is more bursty than Poisson, then \eqref{eq:UB} will still be an upper bound as shown from the results in \cite{squeezing1996}.

The aggregation of the packet arrival processes from the
$\left|{\mathcal{A}}\right|$ nodes of the user  (i.e., $\sum\limits_{i \in {\mathcal{A}}} {{a_i}\left( n \right)}$ in \eqref{eq:queue}) can be modeled as a Poisson process in vehicle communication scenarios as well as other machine type communication scenarios \cite{Mehdi2013Performance,3GPP2012MTC}. For a Poisson arrival process, the effective bandwidth is
\begin{align}
E^{B}(\theta) = \frac{{{\lambda }}}{{T_\mathrm{f}{\theta}}}\left( {{e^{{\theta}}} - 1} \right),\label{eq:EBPoisson}
\end{align}
where  $\lambda$ is the average number of the packets arrived at the queue during one frame, which is identical for all frames.

Substituting \eqref{eq:EBPoisson} into \eqref{eq:delay}, we can obtain that $\theta = \ln \left[\frac{T_\mathrm{f}\ln(1/\varepsilon^q)}{\lambda D^q_{\max}}+1 \right]$, such that \eqref{eq:EBPoisson}  can be rewritten as,
\begin{align}
E^{B}(\theta) = \frac{\ln(1/\varepsilon^q)}{D^q_{\max} \ln \left[\frac{T_\mathrm{f}\ln(1/\varepsilon^q)}{\lambda D^q_{\max}}+1\right]}.\label{eq:EBDepson}
\end{align}
If the number of packets transmitted to the user is a constant among frames that satisfies
\begin{align}
s(n) = T_\mathrm{f}E^B(\theta), \label{eq:QoS}
\end{align}
then the queueing delay requirement $(D^q_{\max}, \varepsilon^q)$ can be ensured and the related departure process in \eqref{eq:bn} becomes
\begin{align}
b(n) = \min\{Q(n),T_\mathrm{f}E^B(\theta)\}. \label{eq:QoSbn}
\end{align}

\subsection{Transmit Power}
In what follows, we show that the required transmit power to guarantee the queueing delay requirement for some packets may become unbounded for any given values of $W$ and $N_\mathrm{t}$. We consider the case where $Q(n) \geq T_\mathrm{f}E^B(\theta)$, then $b(n) = T_\mathrm{f} E^B(\theta)$. If the transmit power can guarantee such a departure rate, then for the other case  where $Q(n) < T_\mathrm{f}E^B(\theta)$, $b(n) < T_\mathrm{f}E^B(\theta)$ can also be supported.

Substituting $s(n)$ in \eqref{eq:sn} and $E^B(\theta)$ in \eqref{eq:EBDepson} into \eqref{eq:QoS}, we can derive the required SNR  $\gamma$ to ensure $(D^q_{\max}, \varepsilon^q)$ and ${\varepsilon^c}$ for all packets using the following equation,
\begin{align}
\ln \left( {1 + \gamma} \right) = \frac{{T_\mathrm{f} u\ln 2\ln \left( {1/\varepsilon ^q} \right)}}{{{\phi}{W}D_{\max }^q\ln \left[ {1 + \frac{T_\mathrm{f}{\ln \left( {1/\varepsilon^q} \right)}}{{D_{\max }^q{\lambda}}}} \right]}} + \sqrt {\frac{V}{{{\phi}{W}}}} f_{\rm G}^{ - 1}\left( {\varepsilon^c} \right). \label{eq:power}
\end{align}

Since the elements of ${\bf{h}}$ whose elements are independent
and identically complex Gaussian distributed with zero mean
and unit variance, the instantaneous channel gain $g$ follows the Wishart distribution \cite{Telatar1995Capacity}, whose probability density function is
$f_g\left( x \right) = \frac{1}{{\left( {{N_\mathrm{t}} - 1} \right)!}}{x^{{N_\mathrm{t}} - 1}}{e^{ - x}}$.
In the considered \emph{quasi-static fading channel}, some packets may experience deep fading with channel gain $g$ that is arbitrarily close to zero. Then, the required transmit power in the corresponding frame to achieve $\gamma$, $P(n) \triangleq \frac{{{N_0}W\gamma }}{{\alpha g}}$, is unbounded.

\subsection{Proactive Packet Dropping Mechanism}
To ensure the reliability ${\varepsilon_\mathrm{D}}$ with a finite transmit power, we introduce a proactive packet dropping mechanism. Denote the maximal transmit power of the BS as $P^{\max}$. We allow some packets in deep fading to be dropped before transmission since the required SNR $\gamma$ cannot be achieved with $P(n) \leq P^{\max }$.

From the approximation in \eqref{eq:sn}, we can obtain the number of packets that can be transmitted with $P (n) = P^{\max }$ as follows:
\begin{align}
s^{\max } \approx \frac{{ {\phi}{W}}}{{u\ln 2}}\left\{ {\ln \left[ {1 + \frac{{{\alpha}P^{\max }{g}}}{{{N_0}{W}}}} \right] - \sqrt {\frac{V}{{{\phi}{W}}}} f_{\rm G}^{ - 1}\left( {{\varepsilon^c}} \right)} \right\}\label{eq:smax}.
\end{align}
When ${g} < \frac{{{N_0}{W}\gamma}}{{{\alpha}P^{\max }}}$, $s^{\max} < T_\mathrm{f} E^B(\theta)$. Since  $b(n) = \min\{Q(n),T_\mathrm{f} E^B(\theta)\}$ should be satisfied to ensure  $(D^q_{\max}, {\varepsilon^q})$, we should discard some packets from the head of the queue. Otherwise, $(D^q_{\max}, {\varepsilon^q})$ cannot be satisfied. Denote the number of packets dropped in the $n$th frame as $b^{d}(n) = \max\{b(n) - s^{\max}, 0 \}$. Then, we have
\begin{align}
b^d\left( n \right) = \left\{ {\begin{array}{*{20}{c}}
{\max \left( {{T_\mathrm{f}}E^B({\theta}) - s^{\max },0} \right),{\rm{if }}\;{Q}\left( n \right) \geq {T_\mathrm{f}}E^B({\theta}),}\\
{\max \left( {{Q}\left( n \right) - s^{\max },0} \right),\;\;\;\;\;\;{\rm{ if}}\;{{{Q}\left( n \right) < {T_\mathrm{f}}E^B({\theta}).}}}
\end{array}} \right. \label{eq:bd}
\end{align}
Similar to the time averaged delivery ratio in \cite{Kumar2009QoS}, we define the time average packet dropping ratio as follows:
\begin{align}
\varepsilon^h \triangleq \mathop {\lim }\limits_{N \to \infty } \frac{{\sum\limits_{n = 1}^N {b^d\left( n \right)} }}{{\sum\limits_{n = 1}^N {\sum\limits_{i \in {\mathcal{A}}} {{a_i}\left( n \right)}} }}= \frac{{\mathbb{E}}[b^d\left( n \right) ]}{\lambda} \label{eq:defineh},
\end{align}
where the second equality is obtained under the assumption that the queueing system is stationary and ergodic, and the average is taken over both channel gain and queue length. To obtain $\varepsilon^h$, we consider an upper bound of $b^d\left( n \right)$ as follows:
\begin{align}
b^{U}\left( n \right) = \left\{ {\begin{array}{*{20}{c}}
{\max \left( {{T_\mathrm{f}}E^B({\theta}) - s^{\max },0} \right),{\rm{if }}\; {Q}\left( n \right) > 0,}\\
{0,\quad\quad\quad\quad\quad\quad\quad\quad\quad\quad\;\;{\rm{ if}}\;{{ {Q}\left( n \right)=0,}}}
\end{array}} \right. \nonumber
\end{align}
When ${Q}\left( n \right) \geq {T_\mathrm{f}}E^B({\theta}) $ or ${Q}\left( n \right) = 0$, $b^{U}\left( n \right) = b^d(n)$. When $0 < {Q}\left( n \right) < {T_\mathrm{f}}E^B({\theta})$, $b^{U}\left( n \right) > b^d(n)$. Then, we can derive an upper bound of ${\mathbb{E}}[b^d\left( n \right) ]$ as follows:
\begin{align}
{{\mathbb{E}}[b^U( n) ]} = \eta {\int_0^{\frac{{{N_0}{W}\gamma}}{{{\alpha}P^{\max }}}} {( {T_\mathrm{f} E^B(\theta) - s^{\max }} ){f_g}( g )dg} }.\nonumber
\end{align}
Upon substituting ${{\mathbb{E}}[b^U( n) ]}$ into \eqref{eq:defineh}, we can derive an upper bound of the packet dropping ratio as follows,
\begin{align}
\varepsilon^h < \int_0^{\frac{{{N_0}{W}\gamma}}{{{\alpha}P^{\max }}}} {\left[ {1 - \frac{{s^{\max }}}{{{T_\mathrm{f}}E^B({\theta})}}} \right]{f_g}\left( g \right)dg} \label{eq:eh},
\end{align}
where $\eta = \Pr\{Q(n)>0\} = {{\mathbb{E}}\{\sum\limits_{i \in {\mathcal{A}}} {{a_i}\left( n \right)}\}}/{{\mathbb{E}}[s(n)]}= \frac{\lambda}{{T_\mathrm{f}}E^B({\theta})}$ is applied.\footnote{Simulation results show that $Q(n) = 0$ or ${Q}\left( n \right) \geq {T_\mathrm{f}}E^B({\theta})$ with more than $90$~\% probability (the results are not provided due to the lack of space). Hence,  in the most cases $b^d( n) = b^U( n)$. This suggests that the upper bound ${{\mathbb{E}}[b^U( n) ]}$ is tight, and hence \eqref{eq:eh} is also tight.}

By substituting $s^{\max }$ in \eqref{eq:smax},  and considering \eqref{eq:EBDepson} and \eqref{eq:power}, we have
\begin{align}
\frac{{{s^{\max }}}}{{{T_\mathrm{f}}{E^B}\left( \theta  \right)}} = \frac{\ln\left[1+\frac{\alpha P^{\max} g}{N_0 W}\right]-\sqrt{\frac{V}{\phi W }}f^{-1}_{\rm G}{(\varepsilon^c)}}{\ln\left(1+\gamma\right)-\sqrt{\frac{V}{\phi W }}f^{-1}_{\rm G}{(\varepsilon^c)}} \label{eq:ratio}.
\end{align}
Note that a packet is dropped only if it is transmitted in deep fading. When $g \to 0$, $V$ in \eqref{eq:dispersion} approaches $0$, and hence \eqref{eq:ratio} can be accurately approximated by
\begin{align}
\frac{{{s^{\max }}}}{{{T_\mathrm{f}}{E^B}\left( \theta  \right)}} \approx \frac{\ln\left[1+\frac{\alpha P^{\max} g}{N_0 W}\right]}{\ln\left(1+\gamma\right)} \approx \frac{\alpha P^{\max} g}{N_0 W \ln\left(1+\gamma\right)} \label{eq:appratio}.
\end{align}

Substituting \eqref{eq:appratio} into \eqref{eq:eh}, we obtain
\begin{align}
{\varepsilon ^h} < \int_0^{\frac{{{N_0}W \gamma}}{{\alpha {P^{\max }}}}} {\left[ {1 - \frac{\alpha P^{\max} g}{N_0 W \ln(1+\gamma)}} \right]{f_g}\left( g \right)dg}  \label{eq:consteh}.
\end{align}

\section{Cross-layer Transmission Design}
In this section, we find the optimal resource allocation policy and packet dropping policy that minimize the required maximal transmit power. We consider the cases that $Q(n)>0$. For the other case $Q(n)=0$, $P(n)=0$.

\subsection{Single-user Scenario}
Since the values of $\varepsilon^q$, $\varepsilon^c$ and $\varepsilon^h$ are related to the resource allocation and packet dropping policy, we use the following framework to determine their optimal combination to ensure the reliability $\varepsilon^q+\varepsilon^c+\varepsilon^h \leq \varepsilon_\mathrm{D}$,
\begin{align}
\mathop {\min }\limits_{\varepsilon ^q,\varepsilon ^c, \varepsilon ^h}& \quad P^{\max}, \label{eq:epsilon3}\\
\text{s.t.} & \quad  \eqref{eq:reliability}, \eqref{eq:power}\;\text{and}\; \eqref{eq:consteh}.\nonumber
\end{align}

From the solution of this problem, we can obtain the power allocation policy $P(n)$ and packet dropping policy $b^d(n)$ that minimizes $P^{\max}$. Specifically, with the values of $\varepsilon ^q$ and $\varepsilon ^c$ as well as $D^q_{\max}$, we can obtain the required SNR $\gamma$ from \eqref{eq:power}. Given the values of $\gamma$ and $\varepsilon ^h$, $P^{\max}$ can be obtained from the right hand side of \eqref{eq:consteh}. The optimal power allocation policy to the TTIs when $Q(n)>0$ is given by:
\begin{align}
{P^*}(n) = \left\{ {\begin{array}{*{20}{c}}
{{P^{\max }},\;\;{\rm{if }}\;g < \frac{{{N_0}W\gamma }}{{\alpha {P^{\max }}}},}\\
{\frac{{{N_0}W\gamma }}{{\alpha g}},\;{\rm{if }}\;g > \frac{{{N_0}W\gamma }}{{\alpha {P^{\max }}}}.}
\end{array}} \right.\label{eq:Pn}
\end{align}
Furthermore, by substituting $P^{\max}$ into $s^{\max}$ in \eqref{eq:bd}, the optimal packet dropping policy is obtained.

In the following, we propose a two-step method to find the optimal solution of problem \eqref{eq:epsilon3}. In the first step, the upper bound of the proactive packet dropping probability is fixed as $\varepsilon_{\rm 0}^h \in (0,\varepsilon_\mathrm{D})$. Given $\varepsilon_{\rm 0}^h$, $P^{\max}$ in \eqref{eq:consteh} increases with $\gamma$. Hence, minimizing $P^{\max}$ is equivalent to minimizing $\gamma$. The optimal values of $\varepsilon^q$ and $\varepsilon^c$ that minimize the required $\gamma$ can be obtained by solving the following problem,
\begin{align}
\mathop {\min }\limits_{\varepsilon ^q,\varepsilon ^c}& \quad \eqref{eq:power}\label{eq:combine2}\\
\text{s.t.} & \quad  \varepsilon^q + \varepsilon^c \leq \varepsilon_\mathrm{D}-\varepsilon_0^h. \label{eq:relia2}\tag{\theequation a}
\end{align}
As proved in Appendix \ref{App:combine}, \eqref{eq:power} is strictly convex in $\varepsilon^q$ and $\varepsilon^c$. As shown in \eqref{eq:dispersion}, when the required SNR $\gamma$ is high, $V \approx 1$ and does not depend on SNR.\footnote{Different from the cases in \eqref{eq:ratio} that packets are dropping, the required SNR for successful transmission is high.} Then, there is a unique optimal solution of $\varepsilon^q$ and $\varepsilon^c$ that minimizes $\gamma$. Denote the minimal SNR obtained from problem \eqref{eq:combine2} as $\gamma^*$. Since the right hand side of \eqref{eq:consteh} decreases with $P^{\max}$, for any given $\varepsilon _0^h$, $P^{\max}$ can be obtained numerically from \eqref{eq:consteh}. Then, we obtain the relation between the minimal $P^{\max}$ and $\varepsilon _0^h$ and denote it as $P^{\max}(\varepsilon _0^h)$.

In the second step, we find the optimal $\varepsilon_0^h \in (0,\varepsilon_\mathrm{D})$ that minimizes $P^{\max}(\varepsilon _0^h)$. Since there is no closed-form expression of $P^{\max}(\varepsilon _0^h)$, the exhaustive search method is needed to obtain the optimal $\varepsilon_0^h$ in general. However, numerical results indicate that $P^{\max}(\varepsilon_0^h)$ first decreases and then increases with $\varepsilon_0^h$ (which are omitted due to the lack of space). With this property, we can find the optimal $\varepsilon_0^h$ and the required transmit power to ensure $\varepsilon_\mathrm{D}$ via binary searching \cite{boyd}.

In summary, we can obtain the optimal solution and the minimal transmit power from the two-step method, denoted as $(\varepsilon^{q^*},\varepsilon^{c^*},\varepsilon^{h^*})$ and $P^{{\max}^*}$, respectively. As proved in Appendix \ref{App:optimal}, $(\varepsilon^{q^*},\varepsilon^{c^*},\varepsilon^{h^*})$ is the global optimal solution of problem \eqref{eq:epsilon3}.

\subsection{Multi-user Scenario}
Denote the total bandwidth as $W^{\max}$, we jointly allocate $W_k$, $P_k(n)$ and $b^d_k(n)$ by optimizing $W_k$, $\varepsilon^q$, $\varepsilon^c$, and $\varepsilon^h$. In multi-user scenario, packet dropping only happens when $\sum\limits_{k = 1}^K {\frac{{{N_0}{W_k}\gamma_k}}{{{\alpha_k}g_k}}} > P^{\max}$. However, since the packet dropping ratio depends on the channel states of all the users, it is hard to derive. To overcome this difficulty, we introduce the maximal transmit power to each user, i.e., when $\frac{{{N_0}{W_k}\gamma_k}}{{{\alpha_k}g_k}} \geq P_k^{\rm th }$ some packets are dropped. Then, the upper bound of the packet dropping ratio of the $k$th user can be expressed as
\begin{align}
{\varepsilon_k^h} = \int_0^{\frac{{{N_0} W_k \gamma_k }}{{\alpha {P_k^{\rm th }}}} } {\left[ {1 - \frac{\alpha_k P_k^{\rm th} g}{N_0 W_k \ln(1+\gamma_k)}} \right]{f_g}\left( g \right)dg}\label{eq:eph}.
\end{align}
The optimization framework in the multi-user scenario is:
\begin{align}
\mathop {\mathop {\min }\limits_{{W_k},\varepsilon _k^q,\varepsilon _k^c,\varepsilon _k^h} }\limits_{k = 1,2,...,K} & P^{\max} = \sum\limits_{k = 1}^K {P_k^{\rm th }}  \label{eq:minPt}\\
\text{s.t.}\;&\sum\limits_k^K W_k  \le W^{\max}, \eqref{eq:reliability}, \eqref{eq:power}\; \text{and} \; \eqref{eq:eph}.  \nonumber
\end{align}
Since the event $\{\sum\limits_{k = 1}^K {\frac{{{N_0}{W_k}\gamma_k}}{{{\alpha_k}g_k}}} > P^{\max}\}$ is a subset of $\bigcup\limits_k {\left\{ {\frac{{{N_0}{W_k}{\gamma _k}}}{{{\alpha _k}{g_k}}} \ge P_k^{{\rm{th}}}} \right\}} $, the packet dropping ratio is overestimated. Hence, $P^{\max}$ obtained from \eqref{eq:minPt} is an upper bound of the minimal transmit power that is required to ensure the QoS. Given ${W_k}$, the transmit power allocation policy among subsequent TTIs and packet dropping policy is similar to that in single-user scenario, i.e., \eqref{eq:Pn} and \eqref{eq:bd}.

Note that the number of symbols transmitted in each DL phase $n_k^s = \phi W_k$ is an integer. Thus, $W_k$, $k=1,...,K$ are discrete variables, and \eqref{eq:minPt} is a mixed-integer programming problem. To reduce the complexity, a heuristic algorithm is proposed in Table I. The basic idea is similar to the steepest descent method \cite{boyd}. Given the values of the discrete variables $W_k$, the problem can be decomposed into $K$ single-user problems similar to \eqref{eq:epsilon3}, which can be solved by the two-step method. We refer to the $K$ single-user problems as \emph{subproblem I}. The bandwidth allocation algorithm includes $\phi W^{\max}-K$ steps. In each step, $1/\phi$ bandwidth is allocated to one of the $K$ users that leads to the steepest total transmit power descent. The proposed algorithm only needs to solve subproblem I $K{(\phi W^{\max}-K)}$ times, and hence the complexity is $O\left(K{(\phi W^{\max}-K)}\right)$.

\renewcommand{\algorithmicrequire}{\textbf{Input:}}
\renewcommand{\algorithmicensure}{\textbf{Output:}}
\begin{table}[htb]\small
\caption{Bandwidth Allocation Algorithm}
\vspace{-0.6cm}
\begin{tabular}{p{8.5cm}}
\\\hline
\end{tabular}
\vspace{-0.4cm}
\begin{algorithmic}[1]
\REQUIRE Number of users $K$, total bandwidth $W^{\max}$, duration of each DL phase $\phi$, packet size $u$, noise spectral density $N_0$, number of transmit antennas $N_\mathrm{t}$, average channel gains of users $\alpha_k$, $k=1,...,K$.
\ENSURE Bandwidth allocation $W^*_k$, $k=1,...,K$.
\STATE Set $n^s_k(0) = 1$, $k=1,...,K$. Set $l = 1$.
\STATE Solve subproblem I with given $W_k(0) = n^s_k(0)/\phi$, and obtain the total transmit power $P^{\max}{(0)}$.
\WHILE{$l \leq \phi W^{\max} - K $}
\STATE Set $\hat{k} = 1$
\WHILE{$\hat{k} \leq K$}
\STATE $n^s_{\hat{k}}{(l)} := n^s_{\hat{k}}(l-1) + 1$; $n^s_{{k}}{(l)} := n^s_{{k}}{(l-1)}$, $k \ne \hat{k}$.
\STATE Solve subproblem I with $W_k{(l)} = n^s_k(l)/\phi$, and obtain $\hat{P}_{\hat{k}}^{\max}{(l)}$.
\STATE $\hat{k} := \hat{k}+1$.
\ENDWHILE
\STATE $k^* := \arg \mathop {\min }\limits_{\hat k} \hat P_{\hat k}^{{\max}}{(l)}$.
\STATE $n^s_{{k}^*}{(l)} := n^s_{{k}^*}{(l-1)} + 1$; $n^s_{{k}}{(l)} := n^s_{{k}}{(l-1)}$, $k \ne {k}^*$.
\STATE $l := l + 1$.
\ENDWHILE
\RETURN $W^*_k = n^s_{{k}}{(l-1)}/\phi, k=1,...,K$.
\end{algorithmic}
\vspace{-0.4cm}
\begin{tabular}{p{8.5cm}}
\\
\hline
\end{tabular}
\vspace{-0.6cm}
\end{table}

\section{Simulation Results}
In this section, we first validate the optimality of the proposed policy. Then, we show the effect of $\varepsilon _k^q$, $\varepsilon _k^c$ and $\varepsilon _k^h$ on the system performance. The three factors that lead to packet loss have not been considered in existing literatures, and hence we do not compare our policy with existing policies.

The users are uniformly distributed with distances from the BS $50$~m $\sim$ $200$~m. The concerned area of each user is a circle region with diameter $d_{\rm c} = 50$~m. The sensors are uniformly distributed with density $0.01$~$\text{user}/\text{m}^2$. Each node uploads packets with rate $10$~packes/s, and each user needs to download packets that are uploaded to the BS from the nodes in the concerned areas of it. Other parameters are listed in Table II, unless otherwise specified.

\begin{table}[htbp]
\vspace{-0.4cm}
\small
\renewcommand{\arraystretch}{1.3}
\caption{Parameters \cite{A2014Scenarios}}
\begin{center}\vspace{-0.2cm}
\begin{tabular}{|p{5cm}|p{2.5cm}|}
  \hline
  Reliability requirement $\varepsilon _\mathrm{D}$ & $1-99.99999\%$ \\\hline
  Queueing delay requirement $D^q_{\max}$ & $0.9$~ms \\\hline
  Duration of each frame (equals to TTI) & $0.1$~ms \\\hline
  Duration of downlink phase & $0.05$~ms \\\hline
  Single-sided noise spectral density $N_0$ & $-173$~dBm/Hz \\\hline
  Packet size $u$ & $20$~bytes\\\hline
  Path loss model $10\lg(\alpha_k)$ & $35.3+37.6 \lg(d_k)$ \\\hline
\end{tabular}
\end{center}
\vspace{-0.4cm}
\end{table}

The required $P^{\max}$ obtained by the proposed algorithm and exhaustive search method are provided in Table III, which illustrate that the proposed algorithm is near-optimal. Because the complexity of exhaustive search method is extremely high with large $W^{\max}$, we only provide results with small values of $W^{\max}$ and $K$. However, the proposed algorithm can be applied to systems with large $W^{\max}$ and $K$, as shown in Table IV.

\begin{table}[btp]
\vspace{-0.4cm}
\small
\renewcommand{\arraystretch}{1.3}
\caption{Required Transmit Power, $W^{\max} = 1$~MHz, $N_\mathrm{t} = 2$}
\begin{center}\vspace{-0.2cm}
\begin{tabular}{|p{2.8cm}|p{1.2cm}|p{1cm}|p{0.8cm}|}
\hline
  Number of users $K$& 2& 4 & 6 \\\hline
  Proposed Algorithm & $0.0216$~W& $0.155$~W& $5.26$~W  \\\hline
  Exhaustive Search & $0.0216$~W& $0.155$~W& $5.26$~W  \\\hline
\end{tabular}
\end{center}
\vspace{-0.2cm}
\end{table}

\begin{table}[btp]
\vspace{-0.4cm}
\small
\renewcommand{\arraystretch}{1.3}
\caption{Required Transmit Power, $K=40$, $N_\mathrm{t} = 2$}
\begin{center}\vspace{-0.2cm}
\begin{tabular}{|p{2.8cm}|p{1.2cm}|p{1cm}|p{0.8cm}|}
\hline
  Bandwidth $W^{\max}$& $6$~MHz & $7$~MHz & $8$~MHz \\\hline
  Proposed Algorithm & $35.7$~W& $6.92$~W& $3.28$~W  \\\hline
\end{tabular}
\end{center}
\vspace{-0.4cm}
\end{table}

\begin{figure}[htbp]
        \vspace{-0.2cm}
        \centering
        \begin{minipage}[t]{0.4\textwidth}
        \includegraphics[width=1\textwidth]{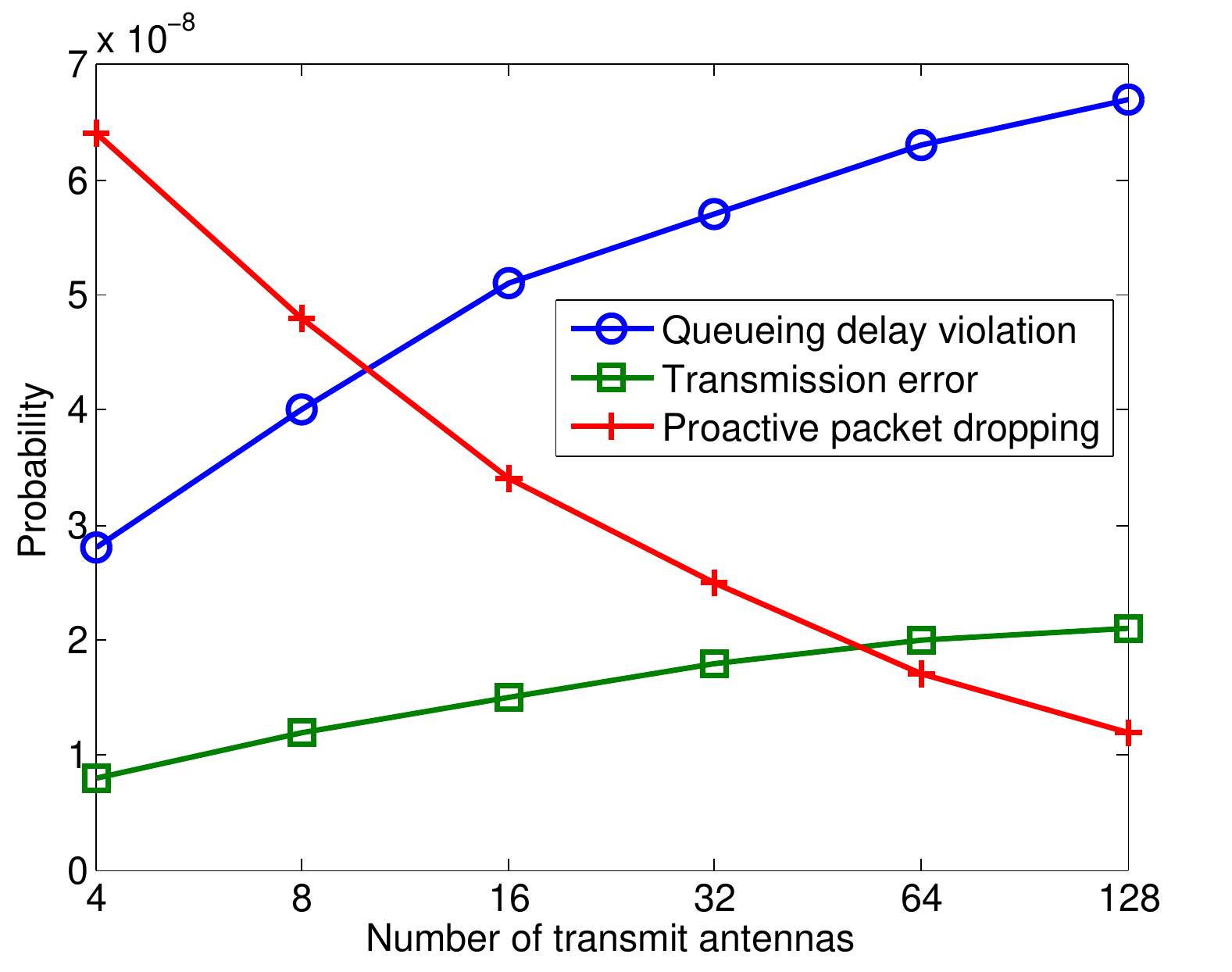}
        \end{minipage}
        \vspace{-0.2cm}
        \caption{$K=1$, the user-BS distance is $200$~m, the bandwidth is $0.5$~Mhz, and the number of nodes in the concerned area is $100$.}
        \label{fig:epsons}
        \vspace{-0.4cm}
\end{figure}
The optimal values of ${\varepsilon _k^c}$, ${\varepsilon _k^q}$ and ${\varepsilon _k^h}$ that minimize the transmit power are illustrated in Fig. \ref{fig:epsons}, which are obtained by \eqref{eq:epsilon3} in single-user scenario. The results show that ${\varepsilon _k^c}$, ${\varepsilon _k^q}$ and ${\varepsilon _k^h}$ are in the same order of magnitude. Hence, when design transmission policy for tactile internet, we conclude that we can not ignore any of these factors.

\section{Conclusion}
In this paper, we showed how to transmit short packets in tactile internet under ultra-low E2E delay and ultra-high reliability requirement. Both queuing delay and transmission delay were considered in the E2E delay, and the transmission error probability, queueing delay violation probability, and packet dropping probability were taken into account in the reliability. The queue state and channel state information dependent transmission policy was optimized to minimize the required maximal transmit power of the BS. Bandwidth allocation was also optimized in multi-user scenario. Simulation results showed that the transmission error probability, queueing delay violation probability, and packet dropping probability are in the same order of magnitude.

\appendices
\section{Proof of the convexity of \eqref{eq:power}}
\label{App:combine}
\renewcommand{\theequation}{A.\arabic{equation}}
\setcounter{equation}{0}
\begin{proof}
The Gaussian Q-function is  ${f_{\rm G}}\left( x \right) = \frac{1}{{\sqrt {2\pi } }}\int_x^\infty  {\exp \left( { - \frac{{{\tau ^2}}}{2}} \right)} d\tau$. Then, ${f'_{\rm G}}\left( x \right) \buildrel \Delta \over =  - \frac{1}{{\sqrt {2\pi } }}{e^{ - {x^2}/2}} < 0$ and ${f''_{\rm G}}\left( x \right) = \frac{x}{{\sqrt {2\pi } }}{e^{ - {x^2}/2}}$, which is positive when $x > 0$, (i.e., ${f_{\rm G}}\left( x \right) < 0.5$). Thus, ${f_{\rm G}}\left( x \right)$ is a decreasing and strictly convex function when $x > 0$. Since $\varepsilon^c < 0.5$, and the inverse function of a decreasing and strictly convex function is also strictly convex \cite{boyd}, $f_{\rm G}^{ - 1}\left( {\varepsilon^c} \right)$ is strictly convex in $\varepsilon^c$. Hence, the second term of \eqref{eq:power} is strictly convex in $\varepsilon^c$.

To prove that the first term of \eqref{eq:power} is strictly convex in $\varepsilon^q$, we derive the second order derivative of it. Denote $y =  - \ln \left( {\varepsilon^q} \right) > 0$ and $z = \frac{T_\mathrm{f}}{{D_{\max }^q{\lambda}}} > 0$. Then, by removing the constants that are not relevant to $\varepsilon^q$, the first term of \eqref{eq:power} can be expressed as follows,
\begin{align}
f\left( y \right) = \frac{{ y}}{{\ln \left( {1 + z y} \right)}}. \label{eq:first}
\end{align}
After some regular derivations, we can obtain that
\begin{align}
\frac{{{d^2}f}}{{d{{\left( {\varepsilon^q} \right)}^2}}} = \frac{f_{\rm num}(y,z) }{{{{\left[ {\ln \left( {1 + zy} \right)} \right]}^3}{{\left( {1 + zy} \right)}^2}{{\left( {\varepsilon^q} \right)}^2}}},\label{eq:finalresult}
\end{align}
where $f_{\rm num}(y,z) = {{\left( {1 + zy} \right)}^2}{{\left[ {\ln \left( {1 + zy} \right)} \right]}^2}+ 2{z^2}y$ $- \left( {2z + zy + {z^2}y + {z^2}{y^2}} \right)\ln \left( {1 + zy} \right)$. It is not hard to prove that $f_{\rm num}(y,z)$ is positive for all $y>16.1$ (i.e., $\varepsilon^q < 10^{-7}$), $z > 0$ (The computing details are omitted due to lack of space). Hence, the first term of \eqref{eq:power} is strictly convex in $\varepsilon^q$. This completes the proof.
\end{proof}

\section{Proof of the optimality of the two-step method}
\label{App:optimal}
\renewcommand{\theequation}{B.\arabic{equation}}
\setcounter{equation}{0}
\begin{proof}
Denote an arbitrary feasible solution of problem \eqref{eq:epsilon3} and the related transmit power as $(\tilde{\varepsilon}^{q},\tilde{\varepsilon}^{c},\tilde{\varepsilon}^{h})$ and $\tilde{P}^{{\max}}$, respectively. Given $\tilde{\varepsilon}^{h}$, we can obtain the minimal transmit power $P^{\max}(\tilde{\varepsilon}^{h}) \leq \tilde{P}^{{\max}}$ via the first step of the two-step method. In the second step, the optimal $\varepsilon^{h^*}$ is obtained such that $P^{\max^*} \leq P^{\max}(\tilde{\varepsilon}^{h})$. Therefore, $P^{\max^*} \leq \tilde{P}^{{\max}}$. The proof follows.
\end{proof}

\bibliographystyle{IEEEtran}
\bibliography{ref}

\end{document}